\documentclass[aps,preprintnumbers,amsmath,amssymb,superscriptaddress,floatfix,nofootinbib]{revtex4}

\usepackage{lipsum}
\usepackage{graphicx}
\usepackage{epsfig}
\usepackage{epstopdf}
\usepackage{hyperref}
\usepackage{amsmath}
\usepackage{amsfonts}
\usepackage{amssymb}

\usepackage{subfig}
\usepackage{slashed}
\usepackage{soul}

\usepackage[normalem]{ulem}  
\usepackage{float}
\usepackage{subfig}

\usepackage{epsfig}

\usepackage{axodraw4j}
\usepackage{pstricks}
\usepackage{color}

\begin{document}

\title{Study of the $DKK$ and $DK \bar K$ systems}

\author{V.~R.~Debastiani}
\email{vinicius.rodrigues@ific.uv.es}
\affiliation{Departamento de
F\'{\i}sica Te\'orica and IFIC, Centro Mixto Universidad de
Valencia-CSIC Institutos de Investigaci\'on de Paterna, Aptdo.
22085, 46071 Valencia, Spain}

\author{J.~M.~Dias}
\email{jorgivan.morais@ific.uv.es}
\affiliation{Departamento de
F\'{\i}sica Te\'orica and IFIC, Centro Mixto Universidad de
Valencia-CSIC Institutos de Investigaci\'on de Paterna, Aptdo.
22085, 46071 Valencia, Spain}
\affiliation{Instituto de F\'{\i}sica, Universidade de S\~{a}o Paulo,
C.P. 66318, 05389-970 S\~{a}o Paulo, SP, Brazil}

\author{E.~Oset}
\email{oset@ific.uv.es}
\affiliation{Departamento de
F\'{\i}sica Te\'orica and IFIC, Centro Mixto Universidad de
Valencia-CSIC Institutos de Investigaci\'on de Paterna, Aptdo.
22085, 46071 Valencia, Spain}

\date{\today}

\begin{abstract}
Using the Fixed Center Approximation to Faddeev equations we
have investigated the $DKK$ and $DK\bar{K}$ three-body systems,
considering that the $DK$ dynamically generates, through its $I=0$
component, the $D^*_{s0}(2317)$ molecule. According to our findings,
for $DK\bar{K}$ interaction we have found an evidence of a state
$I(J^P)=1/2(0^-)$ just above the $D^*_{s0}(2317)\bar{K}$ threshold and around
the $Df_0(980)$ thresholds, with mass about $2833 - 2858$ MeV, made
mostly of $Df_0(980)$. On the other hand, no evidence related to a state
from the $DKK$ interaction is found. The state found could be seen in the
$\pi \pi D$ invariant mass.
\end{abstract}

\maketitle

\section{Introduction}

The study of three-body systems is one of the starting points in the study
of nuclei and nuclear dynamics. The traditional Quantum Mechanical approach
to this problem is based on the Faddeev equations \cite{faddeev} and the main
application was done for three nucleons systems. The simplicity of the Faddeev
equations is deceiving since in practice its  evaluation is very involved and one
approximation or another is done to solve them. One popular choice is the use
of separable potentials to construct the two-body scattering amplitudes via the
Alt-Grassberger-Sandhas (AGS) form of the Faddeev equations \cite{Alt:1967fx}.
Incorporation of chiral symmetry into the scheme has lead to interesting developments
\cite{Epelbaum:2005pn}. Another way to tackle these three-body systems is using
a variational method \cite{Hiyama:2003cu,Dote:2008hw,Kanada-Enyo:2008wsu}.
Gradually, other systems involving not only nucleons or hyperons but mesons were
tackled. The interaction of $K^-d$ at threshold was thoroughly investigated using
Faddeev equations \cite{Toker:1981zh,Torres:1986mr}, or approximations to it, basically
the Fixed Center Approximation (FCA) \cite{Kamalov:2000iy}. The investigation of a
possible state of $K^- pp$ nature has also received much attention
\cite{Dote:2008in,Shevchenko:2006xy,Ikeda:2007nz,Bayar:2011qj,Bayar:2012hn,Uchino:2011jt,bicudo}
and, according to the calculations done in Ref. \cite{sekiramos}, the recent J-PARC
experiment \cite{jparc} has found support for this state.

Another step in this direction was the investigation of systems with two
mesons and one baryon. Surprisingly it was found in Refs.~\cite{alber1,alber2,alber3}
that with such systems one could obtain the low energy baryon states
of $J^P=1/2^+$. Work in this direction with different methods was also
done in Ref. \cite{kbarkbarn} for the $\bar K \bar K N$ system and in Ref.~\cite{kkbarn}
for the $K \bar K N$ system. In this latter case a bound system developed,
giving rise to a $N^*$ state around 1920 MeV, mostly made of a $N a_0(980)$,
which was also predicted in Ref. \cite{alber3}.

Systems of three mesons also followed, and in Ref. \cite{albermeson}
the $\phi K \bar K $ system was studied and shown to reproduce the
properties of the $\phi(2170)$. Similarly, in Ref. \cite{kkkbar} the $K K \bar K $
system is studied and a bound cluster found is associated to the $K(1460)$.
Another similar system, the $\pi K \bar K$  is studied in Ref. \cite{albereta}
and the state found is associated to the $\pi(1300)$. The $\eta K \bar K$ and
$\eta' K \bar K$ systems are also studied in Refs. \cite{albereta,xiaoliang,ollereta}
and they are revised in Ref. \cite{albernew} with the full Faddeev equations
and more solid results.

An important result was found in Refs. \cite{alber1,alber2,alber3,albermeson}.
In the Faddeev equations one uses input from the two-body amplitudes of
the different components and the off-shell part of the amplitudes appears in
the calculations. This off-shell part is unphysical and observables cannot
depend upon it. The finding in those works was that the use of chiral Lagrangians provides three-body contact terms that cancel the off-shell
two-body contributions. In other calculations empirical three-body forces are
introduced which might have some genuine part, but an important part of it
will serve the purpose of effectively cancelling these unphysical off-shell
contributions. Rather than putting these terms empirically, and fitting them
to some data, the message of those works is that to make predictions it is
safer to use as input only on-shell two-body amplitudes, without extra three-body
terms, and an example of it is given in Ref. \cite{alber3}.

Extension to the charm sector was also done. The $DNN$ system,
analogous to the $\bar K NN$ system is studied in Ref. \cite{japocola},
and the $NDK$, $\bar{K} DN$ and $ND\bar{D}$ molecules are studied
in Ref. \cite{Xiao:2011rc}. In the hidden charm sector a resonance is
found for the $J/\psi K \bar K$ system which is associated to the $Y(4260)$
in Ref. \cite{alberdani}. Closer to our work is the one of Ref.~\cite{albermari}
where the $DK\bar{K}$ is studied using QCD sum rules and Faddeev
equations and in both methods a state coupling strongly to $Df_0(980)$ is
found. We will study this system with a different method, and in addition
the $DKK$ system.

The former review of work done shows a constant feature, which is
that systems that add $K \bar K$ to another particle turn out to
generate states in which the $K \bar K$ clusters around the $f_0(980)$
or the $a_0(980)$. The  $DKK$ system benefits from
the $DK$ attraction that forms the $D_{s0}^*(2317)$ according to works
using chiral Lagrangians and unitary approach
\cite{kolo,hofmann,chiang,danids,hanhart1,hanhart2}. It is also supported
by analysis of lattice QCD data \cite{sasa}. However, the $K K$ interaction
is repulsive and the system might not bind. On the other hand, the
$DK \bar K$ system has repulsion for $D \bar K$ in $I=1$, and attraction
for $I=0$, and the $DK$ interaction is attractive, as it is also the $K \bar K$.
Altogether this latter system could have more chances to bind than the
$DKK$ system, but a detailed calculation is called for to find the answer,
and this is the purpose of the present work.

The starting point of our approach is to use the FCA with a preexisting
molecule, which is the $D_{s0}^*(2317)$, formed from the $DK$ interaction.
On top of that, another $K$ (or $\bar K$) is introduced which is allowed to
undergo multiple scattering with the $D$ and $K$ components of the molecule.
The resulting thing, as we shall see, is that in the $DKK$ system we do not
see a signal of a three-body bound state, but in the $DK \bar K$ system we
find a peak which we interpret as the $K \bar K$ fusing to produce the
$f_0(980)$ which gets then bound to the $D$ meson, and a narrow peak
appears at an energy below the $D f_0(980)$ threshold. Such state could
be seen in the $\pi \pi D$ invariant mass.


\section{Formalism}

The Fixed Center Approximation (FCA) to Faddeev
equations is useful when a light hadron $H_3$ interacts
with a cluster $H$ composed of two other hadrons $H_1$ and
$H_2$, $H[H_1\,H_2]$, which are heavier than the first
one, i.e. $M_{(H[H_1\,H_2])}>M_{H_3}$. This cluster comes
out from the two-body interaction between the hadrons $H_1$
and $H_2$ that can be described using a chiral unitary approach
in coupled channels. Hence, the Faddeev equations in this
approximation have as an input the two-body $t$ matrices
for the different pairs of mesons which form the system
and, in this way the generated bound states and resonances are
encoded. In our case, we have $H_1=D$ and $H_2=K$ while
$H_3=\bar{K}$ if we consider the $DK\bar{K}$ interaction
or $H_3=K$ for the $DKK$ system. Both three-body
interactions involve the $D^*_{s0}(2317)$ and $f_0(980)/a_0(980)$
molecules that, according to Refs.~\cite{danids,oller} are
dynamically generated through $DK$ and $K\bar{K}$ interactions,
respectively, taking into account their associated coupled
channels. Therefore, we shall have the following channels
contributing to the three-body interaction systems we are concerned:
(1) $K^-[D^+K^0]$, (2) $K^-[D^0K^+]$, (3) $\bar{K}^0[D^0K^0]$,
(4) $[D^+K^0]K^-$, (5) $[D^0K^+]K^-$ and (6) $[D^0K^0]\bar{K}^0$
for the $DK\bar{K}$ interaction and (1) $K^+[D^+K^0]$, (2) $K^+[D^0K^+]$,
(3) $K^0[D^+K^+]$, (4) $[D^+K^0]K^+$, (5) $[D^0K^+]K^+$ and (6)
$[D^+K^+]K^0$ for $DKK$ system. Note that the states (1), (2) and
(3) are the same as (4), (5) and (6), respectively. Their distinction
is to signify that the interaction in the FCA formalism occurs with the particle
outside the cluster, which is represented by the brackets $[\,.\,.\,.]$, and the
particle of the cluster next to it. This allows for a compact
formulation that describes all the charge exchange steps and
distinguishes the interaction with the right or left component of the
cluster \cite{sekiramos}. These channels will contribute to the
$T_{DK\bar{K}}$ and $T_{DKK}$ three-body scattering matrices and, if
those interactions generate bound states or resonances, they will
manifest as a pole in the solutions of the Faddeev equations. In what
follows we shall discuss how to construct these three-body scattering
matrices and its solution for both, the $DK\bar{K}$ and $DKK$ systems.

\begin{figure}[H]\centering
\includegraphics[width=0.95\textwidth]{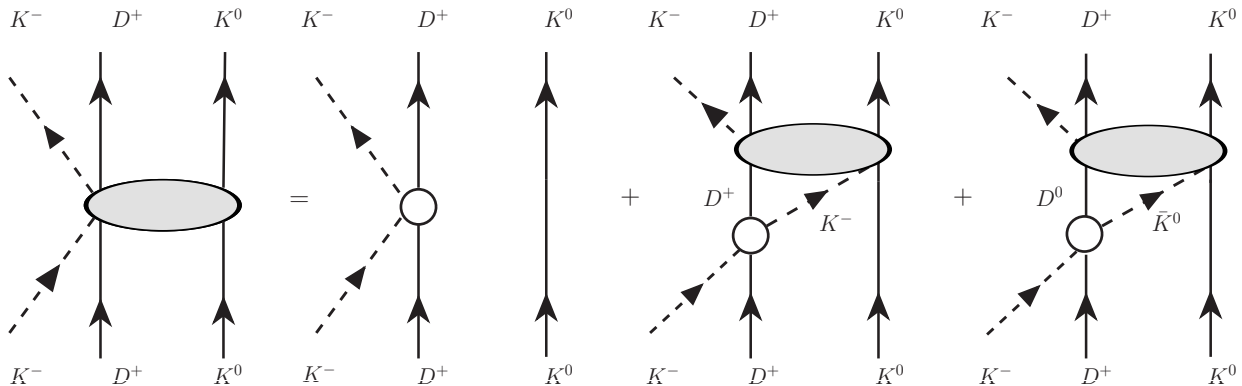}
\caption{Feynman diagrams for the $K^-$ multiple scattering of the process $K^-D^+K^0$.
The white circle indicates the $D\bar{K} \to D\bar{K}$ scattering amplitude
while the gray bubble is associated with the one for $DK\bar{K}$.}\label{diag1}
\end{figure}

\subsection{$DK\bar{K}$ and $DK\bar{K}$ three-body systems}

In order to write the contributions to Faddeev equations of all the channels
mentioned previously, we shall adopt
the following procedure to construct the relevant amplitudes:
for each channel the anti-kaon (kaon) meson to the left side
in (1), (2) and (3) interacts with the hadron to its right
side. Similarly, for the (4), (5) and (6) the $K$ or $\bar{K}$ to the right interacts
with the particle to its left. In doing so, we can distinguish the order of the
anti-kaon (kaon) and two other mesons with which the anti-kaon
(kaon) interacts first and last. This procedure is similar to that
used in Ref.~\cite{sekiramos} to study the $\bar{K}NN$ interaction.
For instance, in the $DK\bar{K}$ system, the channel (1) $K^-[D^+K^0]$
in the initial state, means that the $K^-$ interacts with the
$D^+$ meson to its right. The channel (4) $[D^+ K^0] K^-$ indicates that the $K^-$
interacts with the $K^0$ to its left. This procedure allows us to divide the
multiple anti-kaon (kaon) scattering process in such a way that the formulation
of the multiple scattering becomes easier.

In order to illustrate the structure of the multiple scattering in
the fixed center approximation we define the partition functions
$T^{\rm FCA}_{ij}$, which contain all possible intermediate multiple
steps, where the first index refers to the initial $\bar{K}[DK]$, (1), (2) and (3) or
$[DK]\bar{K}$ (4), (5) and (6) states and the second index to the final state.
If we consider the $K^-[D^+K^0]\to K^-[D^+K^0]$ amplitude
denoted by $T^{\rm FCA}_{11}$, which is diagramatically
represented in Fig.~\ref{diag1}, it provides the following expression \cite{Roca:2010tf,sekiramos}
\begin{eqnarray}\label{t11}
T^{\rm FCA}_{11}(s)=t_1+t_1\,G_0\,T^{\rm FCA}_{41}+t_2\,G_0\,T^{\rm FCA}_{61}\, ,
\end{eqnarray}
which tells us that the transition between the $K^-[D^+K^0]$
to itself is given in terms of a single and double scattering,
coupled to the amplitudes $T^{\rm FCA}_{ij}$ related to the
other channels. As a result, the three-body problem is given in
terms of the $T^{\rm FCA}_{ij}$ partitions, where the $i,\, j$ indices
run from 1 to 6 and stand for the initial and final channels,
respectively, and as we will discuss later, can be displayed in
a matrix form.

In Eq.~\eqref{t11}, $s$ is the Mandelstam variable
that is equal to the squared of the three-body energy system,
while $t_1$ and $t_2$ are, respectively, the $D^+K^-\to D^+K^-$
and $D^+K^-\to D^0\bar{K}^0$ two-body scattering amplitudes
studied in Ref.~\cite{danids}, in which the authors have applied the
chiral unitary approach in coupled channels to investigate the
$D\bar{K}$ and $DK$ two-body interaction. $G_0$ is the kaon
propagator \cite{yamafix} between the particles of the cluster, which
is evaluated through the equation below
\begin{equation}\label{g0}
G_0(s)=\frac{1}{2M_{D^*_{s0}}}\int \frac{d^3{\bf q}}{(2\pi)^3}\frac{F_R({\bf q})}{(q^0)^2-\omega^2_K({\bf q})+i\epsilon}\, ,
\end{equation}
with $\omega^2_K({\bf q})\equiv{\bf q}^2+m^2_K$ and $q^0$ is the energy
carried by kaon meson in the cluster rest frame where $F({\bf q})$ is calculated, which corresponds
to the following expression
\begin{eqnarray}
q^0(s)=\frac{s-m^2_K-M^2_{D^*_{s0}}}{2 M_{D^*_{s0}}}\, .
\end{eqnarray}
In this work, we are using the isospin symmetric masses
such that $m_D$ and $m_K$ are the $D$ and $K$ mesons
average masses, respectively, while $M_{D^*_{s0}}$ is
the $D^*_{s0}$ molecule mass. This molecule dynamics does not
come into play explicitly in our formalism. The information on the
molecule is encoded in the function $F_R({\bf q})$ appearing in
Eq.~\eqref{g0}, the form factor, which is related to the
cluster wave function by a Fourier transform, as discussed
in Refs.~\cite{Roca:2010tf,YamagataSekihara:2010pj}.
According to these works, for the form factor to be
used consistently, the theory that generates the bound
states and resonances (clusters), the chiral unitary approach,
which is developed for scattering amplitudes, has to be extended
to wave functions. This was done in those references for $s$-wave
bound states, $s$-wave resonant states as well as in states
with arbitrary angular momentum \cite{acetiwf}. In our work we need the
form factor expression only for $s$-wave bound states, which
is given by \cite{Roca:2010tf}
\begin{eqnarray}\label{ff}
F_R({\bf q})=\frac{1}{N}\int\limits_{|{\bf p}|,|{\bf p}-{\bf q}| < \Lambda} \, d^3{\bf p}\,\,
\frac{1}{M_{D^*_{s0}}-\omega_D({\bf p})-\omega_K({\bf p})}
\,\frac{1}{M_{D^*_{s0}}-\omega_D({\bf p}-{\bf q})-\omega_K({\bf p}-{\bf q})}\, ,
\end{eqnarray}
where $\omega_D({\bf p})\equiv \sqrt{{\bf p}^2+m^2_D}$
and the normalization factor $N$ is
\begin{eqnarray}\label{norm}
N=\int\limits_{|{\bf p}|<\Lambda}\, d^3{\bf p}\,\Big(\,\frac{1}{M_{D^*_{s0}}-\omega_D({\bf p})
-\omega_K({\bf p})}\Big)^2\, .
\end{eqnarray}
The upper integration limit $\Lambda$ has the same
value of the cut-off used to regularize the loop $DK$,
adjusted in order to get the $D^*_{s0}(2317)$
molecule from the $DK$ interaction.

Analogously to $T^{\rm FCA}_{11}$ expressed in
Eq.~\eqref{t11}, we can calculate all the relevant
multiple scattering amplitudes, the partitions $T^{\rm FCA}_{ij}$,
using similar diagrams like the one in Fig.~\ref{diag1}.
As a result, they can be written as
\begin{eqnarray}\label{multsca}
T^{\rm FCA}_{ij}(s)=V^{\rm FCA}_{ij}(s)+\sum\limits_{l=1}^{6}
 \tilde{V}^{\rm FCA}_{il}(s)\,G_0(s)\,T^{\rm FCA}_{lj}(s)\, ,
\end{eqnarray}
where $V_{ij}$ and $\tilde{V}_{il}$ are the elements
of the matrices below
\begin{equation}
\label{vsdkbar}
  V^{\rm FCA} =
  \left (
  \begin{array}{@{\,}cccccc@{\,}}
    t_{1} & 0 & t_{2} & 0 & 0 & 0 \\
    0 & t_{3} & 0 & 0 & 0 & 0 \\
    t_{2} & 0 & t_{4} & 0 & 0 & 0 \\
    0 & 0 & 0 & t_{5} & 0 & 0 \\
    0 & 0 & 0 & 0 & t_{6} & t_{7} \\
    0 & 0 & 0 & 0 & t_{7} & t_{8} \\
  \end{array}
  \right ) ,
  \quad
  \tilde{V}^{\rm FCA} =
  \left (
  \begin{array}{@{\,}cccccc@{\,}}
    0 & 0 & 0 & t_{1} & 0 & t_{2} \\
    0 & 0 & 0 & 0 & t_{3} & 0 \\
    0 & 0 & 0 & t_{2} & 0 & t_{4} \\
    t_{5} & 0 & 0 & 0 & 0 & 0 \\
    0 & t_{6} & t_{7} & 0 & 0 & 0 \\
    0 & t_{7} & t_{8} & 0 & 0 & 0 \\
  \end{array}
  \right ) .
\end{equation}

Therefore, according to Eq.~\eqref{multsca}, in our case
we can solve the three-body problem in terms of the multiple
scattering amplitudes given by partitions $T^{\rm FCA}_{ij}$,
which contain only the $D\bar{K}$ and $K\bar{K}$ two-body amplitudes.
Thus, for the $DK\bar{K}$ system the
solution of the scattering equation, Eq.~\eqref{multsca}, will be
\begin{eqnarray}\label{tdkbar}
T^{\rm FCA}_{ij}(s)=\sum\limits_{l=1}^{6}\Big[\,1-
\tilde{V}^{\rm FCA}(s)\,G_0(s)\,\Big]^{-1}_{il}\,V_{lj}^{\rm FCA}(s)\, .
\end{eqnarray}

Analogously, for the $DKK$ system, we will have the same
solution as in Eq.~\eqref{tdkbar}. However, in this case, the
$\tilde{V}^{\rm FCA}$ and $V^{\rm FCA}$ matrices, in terms of the $DK$ and $KK$ two-body amplitudes, are now given by
\begin{equation}
\label{vsdk}
  V^{\rm FCA} =
  \left (
  \begin{array}{@{\,}cccccc@{\,}}
    \bar{t}_{1} & 0 & 0 & 0 & 0 & 0 \\
    0 & \bar{t}_{2} & \bar{t}_{3} & 0 & 0 & 0 \\
    0 & \bar{t}_{3} & \bar{t}_{4} & 0 & 0 & 0 \\
    0 & 0 & 0 & \bar{t}_{5} & 0 & \bar{t}_{5} \\
    0 & 0 & 0 & 0 & \bar{t}_{6} & 0 \\
    0 & 0 & 0 & \bar{t}_{5} & 0 & \bar{t}_{5} \\
  \end{array}
  \right ) ,
  \quad
  \tilde{V}^{\rm FCA} =
  \left (
  \begin{array}{@{\,}cccccc@{\,}}
    0 & 0 & 0 & \bar{t}_{1} & 0 & 0 \\
    0 & 0 & 0 & 0 & \bar{t}_{2} & \bar{t}_{3} \\
    0 & 0 & 0 & 0 & \bar{t}_{3} & \bar{t}_{4} \\
    \bar{t}_{5} & 0 & \bar{t}_{5} & 0 & 0 & 0 \\
    0 & \bar{t}_{6} & 0 & 0 & 0 & 0 \\
    \bar{t}_{5} & 0 & \bar{t}_{5} & 0 & 0 & 0 \\
  \end{array}
  \right ) .
\end{equation}
The elements of the matrices in Eqs.~\eqref{vsdkbar} and
\eqref{vsdk}, i.e. $t_1,\,t_2,\,.\,.\,.\,,t_8$ and
$\bar{t}_1,\,.\,.\,.,\,\bar{t}_6$ related to the three-body
interaction $DK\bar{K}$ and $DKK$ systems are the
two-body scattering matrices elements, respectively, given by
\begin{equation}\label{ts}
  \begin{array}{@{\,}cc@{\,}}
    t_{1}=t_{D^+K^-\to D^+K^-}\,; & t_{4}=t_{D^0\bar{K}^0\to D^0\bar{K}^0}\,;  \\
    t_{2}=t_{D^+K^-\to D^0\bar{K}^0}\,; & t_{5}=t_{K^0K^-\to K^0K^-}\,;  \\
    t_{3}=t_{D^0K^-\to D^0K^-}\,; & t_{6}=t_{K^+K^-\to K^+K^-}\,;  \\
  \end{array}
  \quad
  \begin{array}{@{\,}c@{\,}}
    t_7=t_{K^+K^-\to K^0\bar{K}^0}\,;  \\
    t_8=t_{K^0\bar{K}^0\to K^0\bar{K}^0}\, ,\\
  \end{array}
\end{equation}
and \vspace{-10pt}
\begin{equation}\label{tsbar}
  \begin{array}{@{\,}cc@{\,}}
   \bar{t}_{1}=t_{D^+K^+\to D^+K^+}\,; & \bar{t}_{4}=t_{D^+K^0\to D^+K^0}\,;  \\
    \bar{t}_{2}=t_{D^0K^+\to D^0K^+}\,; & \bar{t}_{5}=t_{K^+K^0\to K^+K^0}\,;  \\
    \bar{t}_{3}=t_{D^0K^+\to D^+K^0}\,; & \bar{t}_{6}=t_{K^+K^+\to K^+K^+}\,, \\
  \end{array}
\end{equation}
which we shall discuss in the next subsection.

It is important to mention that, in this work, we are using the
Mandl and Shaw normalization, which has different weight factors
for the particle fields. In order to use these factors in a consistent
manner in our problem, we should take into account how they appear in the single-scattering and double-scattering as
well as in the full amplitude. The detailed calculation on how to do this
can be found in
Refs.~\cite{Roca:2010tf,YamagataSekihara:2010pj,yamafix}.
According to these works, this is done multiplying the two-body
amplitudes by the factor $M_{c}/M_{1(2)}$, where $M_{c}$ is the
cluster mass while $M_{1(2)}$ is associated with the mass of
the hadrons $H_1$ and $H_2$. In our case, we have $M_{c}/M_{D}$
for the two-body amplitudes related to the $D\bar{K}(DK)$ and
$M_{c}/M_{K}$ for the one related to the $K\bar{K}(KK)$ appearing in Eqs.~\eqref{ts}
and \eqref{tsbar}.

Once we solve the Faddeev equations for the systems
we are concerned, we have to write this solution in such
a way that it represents the amplitude of a $\bar{K}\, (K)$
meson interacting with the $D^*_{s0}$ molecule, which is
the $DK$ cluster written into an $I=0$ combination. Taking
into account that
$|DK(I=0)\,\rangle = (1/\sqrt{2})\,|\,D^+K^0+D^0K^+\,\rangle$
(recall $(D^+,-D^0)$ is the isospin doublet), and summing the cases
where the odd $\bar{K}\,(K)$ interacts first to the left (right) of the cluster,
and finishes interacting at the left (right) we obtain the following
combination for both $DK\bar{K}$ and $DKK$ system, 
\begin{eqnarray}\label{eq:Fcluster}
T_{X-D^*_{s0}}&=&\frac{1}{2}\Big( T^{\rm FCA}_{11}
+T^{\rm FCA}_{12}+T^{\rm FCA}_{14}+T^{\rm FCA}_{15}
+T^{\rm FCA}_{21}+T^{\rm FCA}_{22}+T^{\rm FCA}_{24}
+T^{\rm FCA}_{25}+T^{\rm FCA}_{41}
+T^{\rm FCA}_{42}+ T^{\rm FCA}_{44}
+T^{\rm FCA}_{45}\nonumber\\
&+&T^{\rm FCA}_{51}+T^{\rm FCA}_{52}
+T^{\rm FCA}_{54}+T^{\rm FCA}_{55}\Big)\, ,
\end{eqnarray}where $X$ denotes a $\bar{K}$ in the $DK\bar{K}$
case and a $K$ meson for $DKK$ interaction.

\subsection{Two-body amplitudes}

In order to solve the Faddeev equations using the FCA
for the systems we are concerned, we need to
know the two-body scattering amplitudes appearing
in Eqs.~\eqref{ts} and \eqref{tsbar}. They were
studied in Refs.~\cite{danids,oller}. These amplitudes are
calculated using the chiral unitary approach (for a review
see \cite{Oller:2000ma}). In this model, the transition
amplitudes between the different pairs of mesons are
extracted from Lagrangians based on symmetries as
chiral and heavy quark symmetries. Then, they are
unitarized using them as the kernels of the Bethe-Salpeter
equation, which in its on-shell factorization form is given by
\begin{equation}\label{bs}
t=(1-v\,G)^{-1}\,v\, ,
\end{equation}
where $G$ is the two meson loop function and its
expression in dimensional regularization method is
\begin{eqnarray}\label{loop}
G(s_i)&=&\frac{1}{16\pi^2}\Big\{ \alpha_i(\mu)
 + log\frac{m^2_1}{\mu^2}+\frac{m_2^2-m^2_1+s_i}
{2s_i}log \frac{m^2_2}{m^2_1}+\frac{p}{\sqrt{s_i}}
 \Big[\, log(s_i-m^2_2+M^2_1+2p\sqrt{s_i})\nonumber\\
 &-& log (-s_i+m^2_2-m^2_1+2p\sqrt{s_i})
 +log(s_i+m^2_2-m^2_1+2p\sqrt{s_i}) - log(-s_i-m^2_2+M^2_1+2p\sqrt{s_i}) \Big]\Big\}\, ,
\end{eqnarray}
with $m_1$ and $m_2$ standing for the $i$-channel
meson masses in the loop and $p$ is the three-momentum
in the two meson center-of-mass energy, $\sqrt{s_i}$.
In Eq.~\eqref{loop} $\mu$ is a scale fixed a priori and the subtraction
constant $\alpha(\mu)$ is a free parameter. In
Ref.~\cite{danids}, $\mu$ is considered to be equal
to $1500$ MeV for the $D\bar{K}$ system, corresponding
to $\alpha_{D\bar{K}}=-1.15$. On the other hand, since
the amount of $DK$ content in $D^*_{s0}(2317)$ is
about $70\%$ \cite{sasa}, we consider
just one channel, with $\alpha_{DK}=-0.925$, adjusted to
provide the $D^*_{s0}(2317)$ peak, corresponding to a cut-off
value equal to $650$ MeV. This value also has to be used as the
upper limit in the integrals given by Eqs.~\eqref{ff} and \eqref{norm}.
For the $f_0(980)/a_0(980)$ we consider the same channels as Refs.~\cite{xiedai,Dias:2016gou} where
a cut-off equal to $600$ MeV was used to regularize the loops, given by
\begin{equation}\label{Gcutoff}
G(s_l)=\int \frac{d^3{\bf q}}{(2\pi)^3}\frac{\omega_1({\bf q}) + \omega_2({\bf q})}{2\omega_1({\bf q})\omega_2({\bf q})}\frac{1}{(P^0)^2-[\omega_1({\bf q}) + \omega_2({\bf q})]^2+i\epsilon}\, ,
\end{equation}
where $(P^0)^2 = s_l$, the two-body center-of-mass energy squared. The index $l$ stands for the following channels: 1) $\pi^+\pi^-$, 2) $\pi^0\pi^0$, 3) $K^+K^-$, 4) $K^0\bar K^0$, 5) $\eta\eta$ and 6) $\pi\eta$. In each channel $\omega_{1(2)}({\bf q}) = \sqrt{{\bf q}^2 + m_{1(2)}^2}\,$, where $m_{1(2)}$ is the mass of the mesons inside the loop.

In order to get the scattering amplitude for the $KK$
interaction, we follow Ref.~\cite{oller}. First, we
have to find the kernel $v$ to be used in Eq.~\eqref{bs}.
This kernel is the lowest order amplitude describing
the $KK$ interaction and it is calculated using the
chiral Lagrangian
\begin{eqnarray}
\label{lagran}
\mathcal{L}_2=\frac{1}{12\,f_{\pi}^2}\langle\,\, (\partial_{\mu}\Phi\,\Phi
- \Phi\,\partial_{\mu}\Phi)^2+M\Phi^4\,\, \rangle\, ,
\end{eqnarray}
where $\langle\,.\,.\,.\,\rangle$ means the trace
in the flavour space of the $SU(3)$ matrices
appearing in $\Phi$ and $M$ while $f_{\pi}$ is
the pion decay constant. The matrices $\Phi$ and
$M$ are given by
\begin{equation}
\label{Mphi}
  \Phi =
  \left (
  \begin{array}{@{\,}ccc@{\,}}
    \frac{\pi^0}{\sqrt{2}}+\frac{\eta_8}{\sqrt{6}} & \pi^+ & K^+\\
    \pi^- & -\frac{\pi^0}{\sqrt{2}}+\frac{\eta_8}{\sqrt{6}} & K^0\\
    K^- & \bar{K}^0 & -\frac{2\,\eta_8}{\sqrt{6}}\\
  \end{array}
  \right )\, ;
   \quad
     M=
    \left(
    \begin{array}{@{\,}ccc@{\,}}
     m^2_{\pi} & 0 & 0\\
    0 & m^2_{\pi} & 0\\
    0 & 0 & 2\,m^2_{K}-m^2_{\pi}\\
  \end{array}
   \right) \, ,
\end{equation}
where in $M$ we have taken the isospin limit ($m_u=m_d$).
Hence, from Eqs.~\eqref{lagran} and \eqref{Mphi}
we can calculate the tree level amplitudes for $K^+K^0$
and $K^+K^+$, which after projection in $s$-wave read as

\begin{align}
\begin{aligned}
v_{K^+K^0\,,\,K^+K^0 }&=\displaystyle\frac{1}{2 f^2_{\pi}}\Big( s_{KK} - 2 m^2_{K} \Big)\, ;\\
\vspace{15pt}
v_{K^+K^+\,,\,K^+K^+}&=\displaystyle\frac{1}{f^2_{\pi}} \Big( s_{KK} - 2 m^2_{K} \Big)\, ,\\
\end{aligned}
\end{align}
where $s_{KK}$ is the Mandelstam variable
$s$ in the $KK$ center-of-mass frame. From these equations one finds that $v^{I=0}_{KK}=0$ (and $t^{I=0}_{KK}=0$) and taking the unitary normalization appropriate for identical particles $|K^+K^+,I=1\rangle=|K^+K^+\rangle/\sqrt{2}$, we find $v^{I=1}_{KK}=\frac{1}{2}v_{K^+K^+\,,\,K^+K^+}$. The $t$ matrix will be $t^{I=1}_{KK}=(1-v^{I=1}_{KK}\,G_{KK})^{-1}\,v^{I=1}_{KK}$, and then $t^{I=1}_{KK}$ has to be multiplied by two to restore the good normalization. Therefore, using these expressions we obtain the $KK$ scattering amplitudes $\bar{t}_5$ and $\bar{t}_6$ present in Eq.~\eqref{tsbar}
($\bar{t}_6=t^{I=1}_{KK}$, $\bar{t}_5=\frac{1}{2}t^{I=1}_{KK}$, with $t^{I=1}_{KK}$ with the good normalization), where we
have used a cut-off of $600$ MeV to regularize
the $KK$ loops, the same that was used in the $K\bar{K}$ and coupled
channels system. After these considerations we are able to determine all the two-body amplitudes in Eqs.~\eqref{ts} and \eqref{tsbar}.

It is worth mentioning that the arguments of the
partitions $T^{\rm FCA}_{ij}(s)$ and the $t_{i}(s_i)$
two-body amplitudes are different. While the former is
written into the three-body center-of-mass energy
$\sqrt{s}$, the latter is given in the two-body
one. In order to write the $\sqrt{s_i}$'s in terms of $\sqrt{s}$,
we are going to use the same transformations used in
Ref.~\cite{pedro,yamafix}, which are
\begin{eqnarray}\label{eq:prescIDK}
s_{DK(D\bar{K})}=m^2_K+m^2_D+\frac{1}{2M^{2}_{D^*_{s0}}}
(s-m^2_K-M^2_{D^*_{s0}})\,(M^2_{D^*_{s0}}+m^2_D-m^2_K),
\end{eqnarray}
where the subscript $DK(D\bar{K})$ stands for the two-body channels
associated with the energy in the center-of-mass of $DK(D\bar{K})$. Analogously,
for the energy in the $KK(K\bar{K}$) center-of-mass, we have
\begin{eqnarray}\label{eq:prescIKK}
s_{KK(K\bar{K})}=2\,m^2_K+\frac{1}{2M^2_{D^{*}_{s0}}}
(s-m^2_K-M^2_{D^{*}_{s0}})\,(M^2_{D^*_{s0}}+m^2_K-m^2_D)\, .
\end{eqnarray}
In this work, we are going to call this set of transformations
Prescription I. 
In order to estimate the uncertainties in our calculations, we will also use another set of transformations, which we are going call Prescription II, given by
\begin{eqnarray}\label{eq:prescIIDK}
s_{DK(D\bar{K})}=\Big( \frac{\sqrt{s}}{M_{D^*_{s0}}+m_K}\Big)^2
\, \Big(m_K+\frac{m_D\,M_{D^*_{s0}}}{m_D+m_K} \Big)^2
-{\bf P}^2_2\, ,
\end{eqnarray}
and
\begin{eqnarray}\label{eq:prescIIKK}
s_{KK(K\bar{K})}=\Big( \frac{\sqrt{s}}{M_{D^*_{s0}}+m_K}\Big)^2
\, \Big(m_K+\frac{m_K\,M_{D^*_{s0}}}{m_D+m_K} \Big)^2
-{\bf P}^2_1\, ,
\end{eqnarray}
where ${\bf P}_1$ and ${\bf P}_2$ stand for the momenta of the $D$ and $K$ mesons in the
cluster, which we take equal and such that the kinetic energy in the $DK$ cluster is of the
order of the binding energy, hence ${\bf P}^2_1={\bf P}^2_2= 2\tilde{\mu}B_{D^*_{s0}}=2\tilde{\mu}\,(m_D+m_K-M_{D^*_{s0}})$, with
$\tilde{\mu}$ the reduced mass of $DK$.
This prescription is based on another one discussed in Refs.~\cite{pedro,bayarren}, which shares the binding energy among the three-particles proportionally to their respective masses.


\section{Results}

\begin{figure}[h!]
 \centering
 \includegraphics[width=0.7\textwidth]{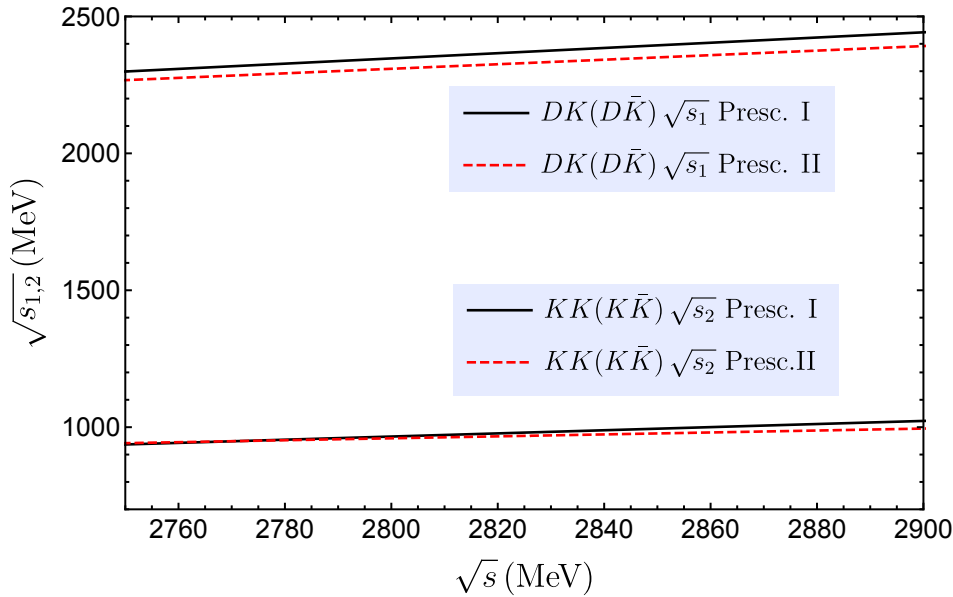}
 \caption{Energy distribution in the center-of-mass of each two-body
 system as a function of the total energy of the three-body system,
 using prescriptions I and II. Here
$s_1 =s_{DK(D \bar K)}$ and $s_2 = s_{KK(K \bar K)}$.
The lower curves are for $KK$ or $K\bar{K}$ and the upper curves are for $DK$ or $D\bar{K}$.}
 \label{fig:energy}
\end{figure}

In all our calculations we use $m_K = 495$ MeV, $m_D = 1865$ MeV,
$m_{D_{s0}^*(2317)} = 2317$ MeV, $m_\pi = 138$ MeV, $m_\eta = 548$
MeV and $f_\pi = 93$ MeV. In Fig.~\ref{fig:energy} we plot the energies in
the center-of-mass of each of the two-body systems as a function of the
energy of the center-of-mass of the three-body system, according to
Eqs.~\eqref{eq:prescIDK}, \eqref{eq:prescIKK}, \eqref{eq:prescIIDK} and
\eqref{eq:prescIIKK}. Both prescriptions map the energy range around 2812 MeV, the threshold
of $D_{s0}^*(2317) K$ (or $D_{s0}^*(2317) \bar K$), to an energy
range around each of the thresholds of the two-body interactions, \textit{i. e.}
the $K K$ system (or $K \bar K $) interact in the energy range around 990
MeV in their center-of-mass, which corresponds to $2 \, m_K$, and the $D K$
(or $D \bar K$) interact in the energy range around 2360 MeV, which
corresponds to $m_K + m_D$.

The main uncertainty in our calculation is the difference between these
two ways of mapping the total energy into the center-of-mass of each
two-body system. This feature was also found in other works using FCA, for instance in Ref. \cite{pedro}. 


\subsection{The $D K \bar K$ system}

\begin{figure}[h!]
  \subfloat[]{%
    \includegraphics[width=0.48\textwidth]{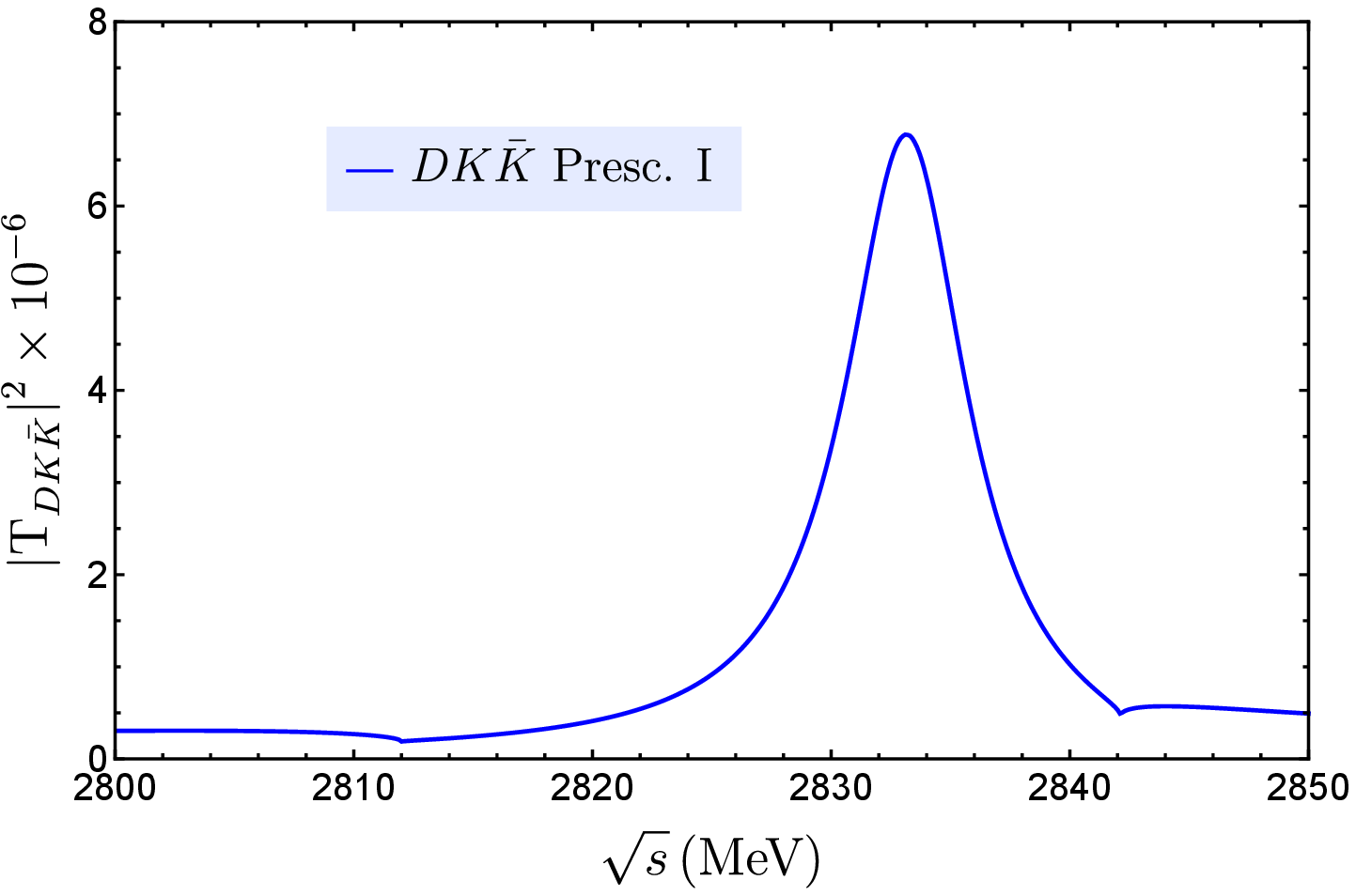} \label{fig:T2DKKbarP1}
  }
  \quad
  \subfloat[]{%
    \includegraphics[width=0.48\textwidth]{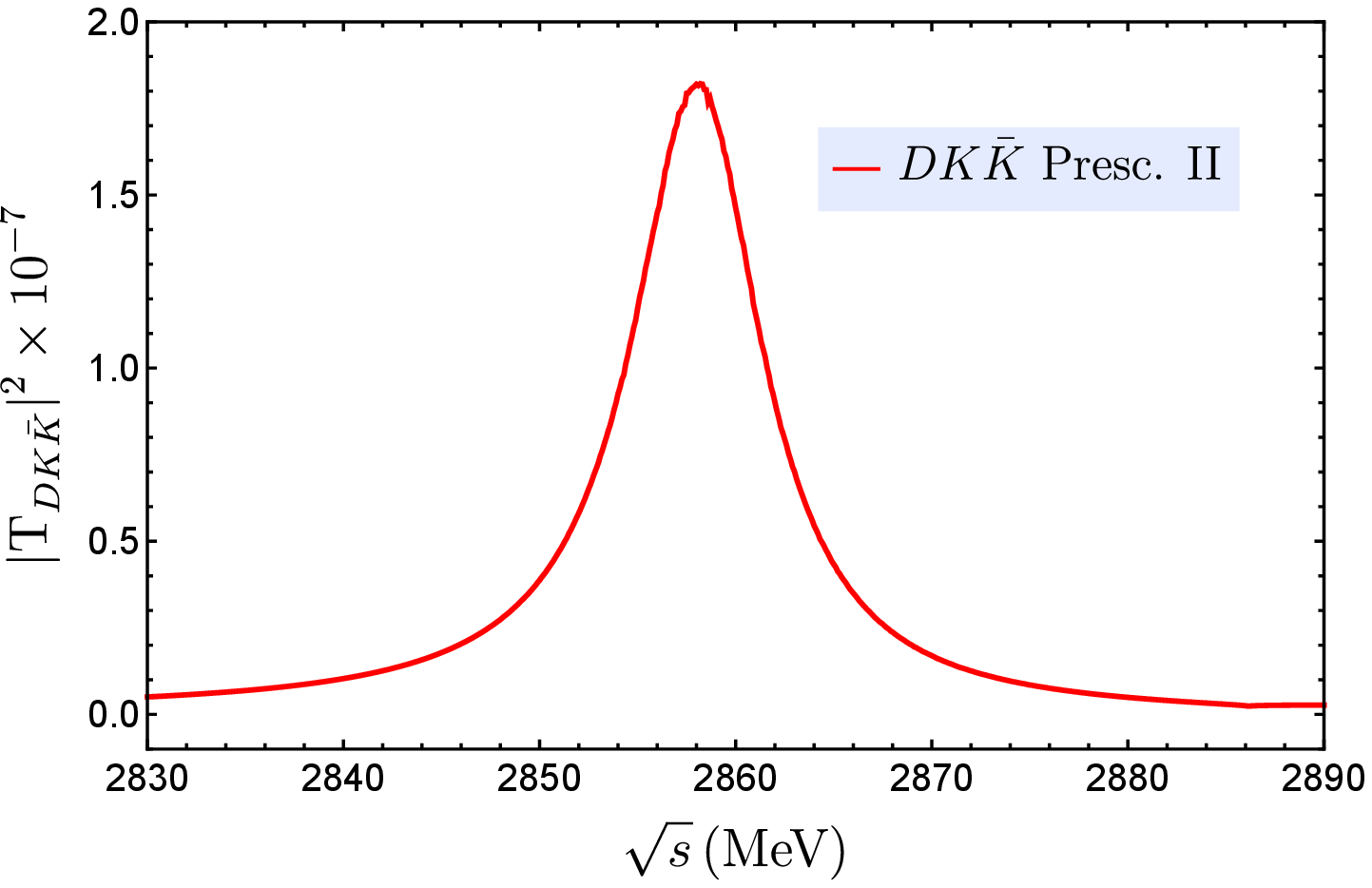} \label{fig:T2DKKbarP2}
  }
  \caption{Results for the total $D K \bar K$ amplitude squared
 using prescriptions I (left) and II (right).}
\end{figure}

In Fig.~\ref{fig:T2DKKbarP1} we show the result of the total Faddeev amplitude squared from Eq.~\eqref{eq:Fcluster} using Prescription I. We can see a strong peak around 2833 MeV, which could be interpreted as a $D [f_0(980)/a_0(980)]$ bound state, since it is below the $D [f_0(980)/a_0(980)]$ threshold of $2855$ MeV. On the other hand, using Prescription II we observe a peak around 2858 MeV, as can seen in Fig.~\ref{fig:T2DKKbarP2}, and now could be interpreted as a $D [f_0(980)/a_0(980)]$ resonance since it is above its threshold.

\begin{figure}[h!]
 \centering
\includegraphics[width=0.7\textwidth]{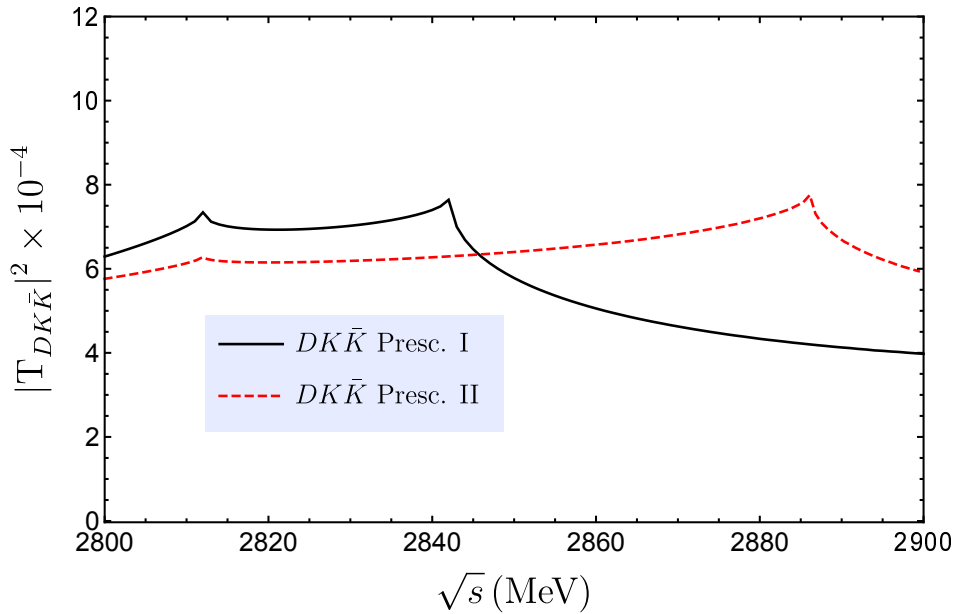}
\caption{Results for the $D K \bar K$ amplitude squared after removing the
$f_0(980)$ contribution, using prescriptions I and II.}
\label{fig:T2nof0}
\end{figure}

In order to investigate if this strong peak in the $D K \bar K$ system
comes mostly from $K \bar K$ merging into $a_0(980)$ or $f_0(980)$,
we have separated the $K \bar K$ amplitudes (that enter in the Faddeev
equations) in isospin basis and selected only one contribution at a time.
In Fig.~\ref{fig:T2nof0} we show the results where the $I=0$ component of $K \bar K$ was removed, therefore there is no $f_0(980)$ contribution. In this figure we can see clearly the shape of the $a_0(980)$ in the three-body amplitude, that peaks around 2842 MeV in Prescription I (and 2886 MeV in Prescription II), which according to Fig.~\ref{fig:energy}, correspond to 990 MeV in the $K \bar K$ center-of-mass, exactly where the $a_0(980)$ peak results from the $I=1$ $K \bar K$ two-body amplitude. Notice that when we removed the $I=0$ isospin component from the $K \bar K$ amplitude the strength of the peaks in $|T_{DD\bar{K}}|^2$ have decreased more than two orders of magnitude in both prescriptions, pointing out that the $f_0(980)$ is indeed the most important contribution coming from $K \bar K$. It is interesting to recall that the same conclusion was obtained in \cite{albermari}, where no apparent signal for $D a_0(980)$ was found. Furthermore, the small cusps seen in both prescriptions at 2812 MeV in Fig.~\ref{fig:T2nof0} correspond to the $D^*_{s0}(2317)\bar{K}$ threshold. In Table \ref{tab:Results} we compile the results of both prescriptions.

 The results for the $D K \bar K$ system point out to the formation of a three-body state: the $D [f_0(980)/a_0(980)]$, in which the $D f_0(980)$ is the strongest contribution in both prescriptions. Specifically, in Prescription I the $Df_0(980)$ state would be bound by about $20$ MeV, while in Prescription II it would correspond to a resonance. This latter result would be similar to the findings of Ref.~\cite{albermari} where a peak is seen at higher energy, forming a $D f_0(980)$ resonant state at 2890 MeV.

As mentioned previously, the difference between the results of prescriptions I and II should be interpreted as the main uncertainty in our approach, but what emerges from both pictures is that a $Df_0(980)$ state is formed, slightly bound or unbound. \vspace{-10pt}
\begin{table}[H]
\caption{\label{tab:Results} Comparison between position and intensity of the
peaks found in the $D K \bar K$ amplitude.}
\begin{center}
\begin{tabular}{| c | c | c | c | c |}
\hline
\rule[-1ex]{0pt}{2.5ex}  & \multicolumn{4}{ c |}{Prescription I      ~~                    Prescription II} \\
\hline
\rule[-1ex]{0pt}{2.5ex}  & $\sqrt{s}$ & $|T|^2$ & $\sqrt{s}$ & $|T|^2$ \\
\hline
\rule[-1ex]{0pt}{2.5ex} Total & $2833$ & $6.8\,\times 10^6$ & $2858$ & $1.8\,\times\,10^7$ \\
\hline
\rule[-1ex]{0pt}{2.5ex} $I=1$ only & $2842$ & $7.7\,\times\, 10^4$ & $2886$ & $7.8\,\times\,10^4$ \\
\hline
\end{tabular}

\end{center}
\end{table}

We would like to note that the theoretical uncertainty of the present method is of the order of $25$ MeV. To put this number in a proper context we can recall that the uncertainty in the QCD sum rules method in Ref.~\cite{albermari} is far larger, with a mass given by $m_{Df_0}=(2926 \pm 237)$ MeV (the uncertainty for the mass in the Faddeev method of Ref.~\cite{albermari} is not given).

\subsection{The $D K K$ system}

\begin{figure}[H]
 \centering
 \includegraphics[width=0.7\textwidth]{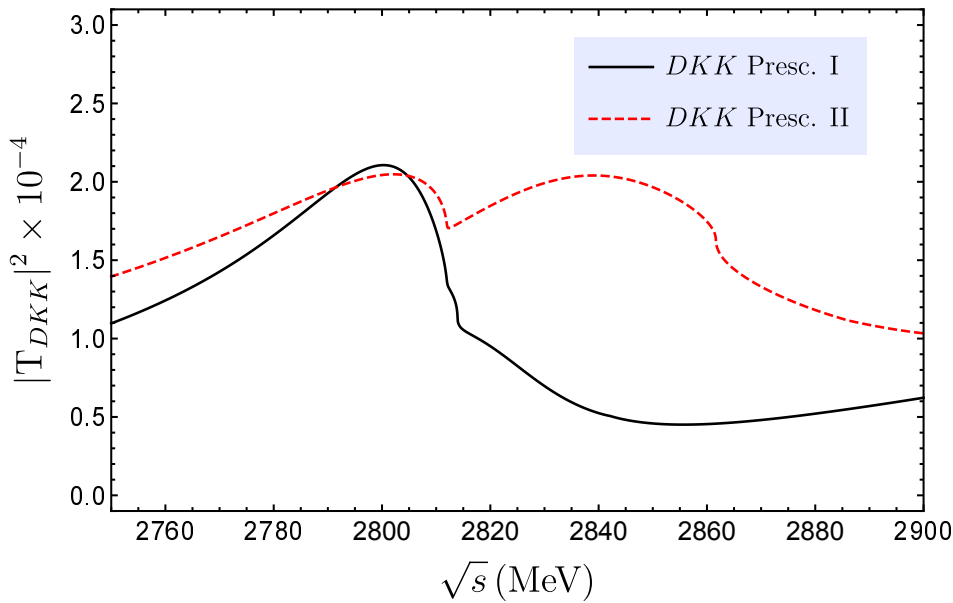}
 \caption{Results for the total $D K K$ amplitude squared using prescriptions I and II.}
 \label{fig:T2DKK}
\end{figure}

In Fig. \ref{fig:T2DKK} we show the $D K K$ total amplitude squared from
Eq.~\eqref{eq:Fcluster} using prescriptions I and II. We can see that in
both prescriptions the amplitude decreases around 2812 MeV which corresponds
to the $D_{s0}^*(2317) K$ threshold, and both have a maximum below this threshold,
while Prescription II also develops a broad structure above threshold, but no clear peak which could indicate that a bound state or a
resonance is found. 

As a physical interpretation we could say that, even though the $D K$ interaction
is attractive and responsible for the strong binding that generates the $D_{s0}^*(2317)$,
the repulsion between $K K$ seems to be of the same magnitude and prevents the
$D K K$ system to form a bound state.

One might be tempted to associate the peak below threshold to a physical state, but this is not the case. Indeed, one should note that the strength of $|T_{DKK}|^2$ in Fig.~\ref{fig:T2DKK} is about three orders of magnitude smaller than for $|T_{DK\bar{K}}|^2$ in Fig.~\ref{fig:T2DKKbarP2}, which simply indicates that no special hadron structure has been formed in this case.

\section{Conclusions}

In this work, we have used the FCA to Faddeev equations in order
to look for bound states or resonances generated from $DK\bar{K}$ and
$DKK$ three-body interactions. The cluster $DK$ in the $I=0$ component is the well known
$D^*_{s0}(2317)$ bound state studied by means of the chiral unitary approach.
From the $DK\bar{K}$ interaction we found a $I(J^P)=1/2(0^-)$ state
with mass about $2833-2858$ MeV, where the uncertainties were estimated
taking into account two different prescriptions to obtain $\sqrt{s_{DK}}$ and
$\sqrt{s_{KK}}$ from the total energy of the system $\sqrt{s}$. Our findings corroborated
those of Ref.~\cite{albermari}, where the authors studied the $DK\bar{K}$
interaction using two different nonperturbative calculation tools, the QCD sum
rules and the Faddeev equations without FCA. They found a state around $2890$ MeV which
is above the $Df_0(980)$ threshold. As we have pointed out before, this state
could be seen in the $\pi\,\pi\, D$ invariant mass distribution. Therefore, as in
Ref.~\cite{albermari}, we also suggest the search for such a state in future
experiments. On the other hand, for the $DKK$ system we found an enhancement
effect, but with a very small strength compared to the $DK\bar{K}$ system and should not be related to a physical bound state. In this case, the repulsion between $K K$ seems to be of the same magnitude as the attraction on the $DK$ interaction, preventing the formation of the three-body molecular state.

\section*{Acknowledgments}

V.~R.~Debastiani wishes to acknowledge the support from the
Programa Santiago Grisolia of Generalitat Valenciana (Exp. GRISOLIA/2015/005).
J.~M.~Dias would like to thank the Brazilian funding agency FAPESP for the financial support.
This work is also partly supported by the Spanish Ministerio de Economia
y Competitividad and European FEDER funds under the contract number
FIS2014-57026-REDT, FIS2014-51948-C2-1-P, and FIS2014-51948-C2-2-P, and
the Generalitat Valenciana in the program Prometeo II-2014/068.

\clearpage

\bibliographystyle{plain}

\end{document}